\documentclass[a4paper, titlepage, 10pt, usenames, table, final]{article}

\usepackage{xr}
\usepackage{geometry}
\usepackage{setspace}

\usepackage[utf8]{inputenc}

\usepackage{graphicx} 
\graphicspath{{./figures/}}
\usepackage{amsmath}
\usepackage{amsfonts}
\DeclareGraphicsExtensions{.png,.pdf,.svg}

\usepackage[hyperref=false,
            url=false,
            isbn=false,
            doi=false,
            eprint=false,
            sorting=none,
            firstinits=true,        
            maxbibnames=99,
            backend=bibtex]{biblatex}
\AtEveryBibitem
{
    \clearfield{number}
    \clearfield{issue}
    \clearfield{eid}
    \clearfield{note}
}
\DefineBibliographyExtras{english}{}
\DefineBibliographyExtras{english}{}
\renewbibmacro{in:}{}
\bibliography{bibliography}

\newcommand{\fref}[1]{Fig.~\ref{#1}}

\newcommand{\eref}[1]{Eq.~\ref{#1}}

\usepackage{caption}
\usepackage{subcaption}

\usepackage[version=3]{mhchem}

\usepackage{verbatim}

\usepackage{siunitx}
\DeclareSIUnit\invcm{\per\centi\metre}
\DeclareSIUnit\meva{\milli\eV\per\angstrom}
\DeclareSIUnit\field{\volt\per\centi\metre}
\DeclareSIUnit\vpa{\volt\per\angstrom}
\DeclareSIUnit\vpa{\volt\per\angstrom}
\newcommand{\waveno}{\widetilde{\nu}}

\renewcommand{\vec}[1]{\mathbf{#1}}

\usepackage{color}
\newcommand{\revchange}[1]{#1}

\title{Molecular dynamics simulation with finite electric fields using Perturbed Neural Network Potentials}

\usepackage{authblk}
\usepackage[symbol*]{footmisc}

\author[1]{Kit Joll}
\author[1,2]{Philipp Schienbein\footnote{to whom correspondence should be addressed, email: p.schienbein@ucl.ac.uk}}
\author[3]{Kevin M. Rosso}
\author[1]{Jochen Blumberger\footnote{to whom correspondence should be addressed, email: j.blumberger@ucl.ac.uk}}  

\affil[1]{Department of Physics and Astronomy and Thomas Young Centre, University College London, London, WC1E 6BT, United Kingdom}
\affil[2]{Department of Physics, Imperial College London, Exhibition Rd, South Kensington, London, SW7 2AZ, United Kingdom}
\affil[3]{Pacific Northwest National Laboratory, Richland, Washington 99354, United States}

\date{ }

\begin{document}

\maketitle

\begin{abstract}
The interaction of condensed phase systems with external electric fields is of major importance in myriad processes 
in nature and technology ranging from the field-directed motion of cells (galvanotaxis), to geochemistry and the 
formation of ice phases on planets, to field-directed chemical catalysis, to energy storage 
and conversion systems including supercapacitors, batteries and solar cells.  
Molecular simulation in the presence of electric fields would give important atomistic insight into these processes 
but applications of the most accurate methods such as ab-initio molecular dynamics are limited in scope by their computational 
expense. Here we introduce Perturbed Neural Network Potential Molecular Dynamics 
(PNNP MD) to push back the accessible time and length scales of such simulations at virtually no loss in accuracy when compared to 
ab-initio molecular dynamics. The total forces on the atoms are expressed in terms of the unperturbed potential 
energy surface represented by a standard neural network potential 
    \revchange{and a field-induced perturbation}
obtained from a 
\revchange{series} 
expansion of the field interaction truncated at first order.
    \revchange{The latter is}
represented in terms of an equivariant graph neural network,
    \revchange{trained on the atomic polar tensor.}
PNNP MD is shown to give excellent results 
for the dielectric relaxation dynamics, the dielectric constant and the field-dependent IR spectrum of liquid water 
    when compared to ab-initio molecular dynamics or experiment, up to surprisingly high field strengths of about $\SI{0.2}{\vpa}$. 
This is remarkable because, in contrast to most previous approaches, the two neural networks on which PNNL MD 
is based are exclusively trained on zero-field molecular configurations demonstrating that the networks 
not only interpolate but also reliably extrapolate the field response. 
    PNNP MD is 
based on rigorous theory yet it is simple, general, modular, and systematically improvable allowing us to obtain 
atomistic insight into the interaction of a wide range of condensed phase systems with 
external electric fields. 
\end{abstract}

\section*{Introduction}

Electric fields are omnipresent in nature and technology. Their presence guides bumblebees to find nectar~\cite{Clarke-2013-Science}, 
and they are believed to cause a superionic ice VII phase on Venus~\cite{Futera-2020-SciAdv}. They play a central role in
myriad electronics and energy conversion devices including field effect transistors, (super-)capacitors, batteries and solar cells.  
In chemistry, electric fields can be used to steer selectivities in catalysis~\cite{BesaluSala-2021-ACSCatal} and to control reactivities~\cite{Shaik-2020-JACS},
whilst in physics they are used to accelerate particles close to the speed of light.  
The field strengths in these examples span an extraordinarily large range.
Atmospheric fields in fair weather are on the order of $10^{-6}\;\si{\vpa}$~\cite{Rycroft-2000-JAtmos} and
electric fields acting as floral cues can be as large as $10^{-5}\;\si{\vpa}$~\cite{Clarke-2013-Science}.
In electrical components the field strength can vary significantly depending on the actual design and the envisaged 
application from about $10^{-10}$ to $10^{-3}\; \si{\vpa}$. Particle accelerators typically operate at a field strength 
of up to $5 \times 10^{-3}\; \si{\vpa}$.  At charged electrodes field strengths on the order of \SI{0.1}{\vpa}~\cite{Toney-1994-Nature} can be found, 
which also marks the onset of chemical bond activation~\cite{Hao-2022-NatCommun}.
The making and breaking of chemical bonds may require more than \SI{1}{\vpa}~\cite{Schirmer-2010-ChemCommun, Ashton-2020-PRL}.

Molecular dynamics~(MD) or Monte Carlo~(MC) simulations are the most suitable methods 
to understand and predict the interaction of liquids, liquid electrolytes and liquid/solid interfaces with external electric fields 
because they sample the correct equilibrium distribution of molecular configurations. Many 
\revchange{finite electric field simulations~\cite{Stengel-2009-Nature}} have been carried out 
with classical force fields~\cite{Zhang-2016-PRB, Sayer-2017-JCP, cox&sprik}
which typically give a good description for pure solvents but 
tend to perform poorly on more complex systems where charge transfer and 
polarization effects become important, e.g., electrode/electrolyte interfaces. 
Ab-initio MD (AIMD) simulations\cite{Marx-Book-2009} solve the electronic structure of the system from first principles 
at every MD time step (usually at the density functional theory~(DFT) level of theory) and thus give the most accurate description for such cases.  
Indeed, AIMD simulations with finite electric fields have been carried out on a number of systems\cite{english2015perspectives}, 
ranging from 
\revchange{crystals~\cite{Stengel-2009-Nature} and }
pure liquid water\cite{Elgabarty-2019-SciRep, Elgabarty-2020-SciAdv} to 
electrode/electrolyte interfaces\cite{zhang2020modelling,capacitance_jun_cheng,zhangsprik_dielectric, Futera-2021-JPCL}. 
A serious disadvantage of AIMD simulations is that they are still computationally extremely demanding, with simulation times
typically limited to a few 10 picoseconds and system sizes limited to a few 100 atoms depending on the density functional chosen. 
As such, AIMD simulations of important field-induced phenomena, such as dielectric relaxation, ionic conductivity\cite{cox&sprik},  
electric double layer formation\cite{Huang-2023-JPCL} or capacitive charging are prohibitively expensive due to the large number 
of atoms required to faithfully model such processes. 

The advent of machine learning (ML) has transformed the field of MD simulations.
Training a ML potential (MLP) from a few hundred explicit electronic structure calculations\cite{Behler-2007-PRL,bartok2010gaussian,Behler-2021-ChemRev}, 
it is now possible to carry out MLMD simulation at AIMD accuracy on truly nanoscale systems over nanosecond time scales.  
Originally, these ML models were designed to calculate potential energies and forces only.
Over the last years, a full bouquet of ML models have been published to predict dipole moments and 
polarizabilities~\cite{Zhang-2020-PRB, Sommers-2020-PCCP,Schuett-2021-ProcMachineLearningRes, Wilkins-2019-PNAS, Kapil-2020-JCP, Shepherd-2021-JPCL, Beckmann-2022-JCTC}, 
as well as other response properties including atomic polar tensors (APT)\cite{Schienbein-2023-JCTC}. 
Following these developments, some ML models have been introduced which explicitly describe the interaction of a molecular system with external 
electric fields\cite{Christensen-2019-JCP, Gastegger-2021-ChemSci, Gao-2022-NatCommun, Shao-2022-ElectronicStruct, Zhang-2023-NatCommun}. 
In most of these approaches the field-dependence of the potential energy surface is 
\revchange{explicitly or implicitly part of the}
MLP~\cite{Christensen-2019-JCP, Gastegger-2021-ChemSci, Gao-2022-NatCommun, Zhang-2023-NatCommun}. Whilst these approaches could 
be successfully applied to molecular systems and liquids, 
\revchange{the electric field is an input parameter to the MLP and therefore training data for different field strengths is required}.

Herein, we introduce a simple,
\revchange{complementary} 
approach for MLMD with electric fields that does not require 
\revchange{the generation of any training data with an electric fields}.
We start from 
\revchange{a standard}
potential energy surface and account for 
the interaction with the electric field 
\revchange{in a perturbative manner by a series expansion}
truncated at first order, noting that the first order term can be written in terms 
of the APT~\cite{Person-1974-JCP, Schienbein-2023-JCTC}. Two MLPs are trained: one standard MLP  
for the unperturbed potential energy surface (here, a committee neural network potential (c-NNP)~\cite{Behler-2007-PRL, Schran-2020-JCP, Behler-2021-ChemRev}) and one ML model 
for the APT (here, an equivariant graph neural network denoted ``APTNN''~\cite{Schienbein-2023-JCTC}). 
Combined 
they form a ``perturbed MLP'' (here, ``perturbed neural network potential'' (PNNP)).  The two networks have no knowledge of electric field effects 
\revchange{(neither explicitly, nor implicitly)}
because they are both trained solely on data and 
\revchange{atomic} 
configurations obtained from zero-field calculations. The interaction with the field 
during MD simulation simulation is 
\revchange{therefore entirely due}
to the first order term in the 
\revchange{series} expansion 
\revchange{given by} 
the product of the APT represented by the 
graph neural network and the external field. 

A PNNP is based on rigorous physical principles and follows precisely the treatment of electric fields in quantum chemical calculations, 
where the total energy is 
\revchange{usually} 
expanded at zero field~\cite{SzaboOstlund}. The accuracy can thus be systematically improved by adding higher order 
contributions, such as polarizabilities and hyperpolarizabilities, in ML representation. Since we do not introduce or modify ML models, the advantages 
and accuracies, but also the limitations of the two employed ML models are inherited. 
Importantly, a PNNP follows the spirit of the modern theory of polarization~\cite{resta_MTP_1992} 
because the APT relates to a \emph{change} of polarization and is thus not affected by the multivaluedness of the polarization in periodic boundary conditions.
This makes PNNP applicable to a broad range of condensed phase systems typically modelled under periodic boundary conditions, 
including solids, liquids, and interfaces. 

In the following section we describe the PNNP approach in detail. After validation we apply the method to simulate the dielectric response 
of pure liquid water. We demonstrate reversible polarization and depolarization of liquid water as the field strength is stepped up and down.  
Then the dielectric relaxation dynamics is analysed in detail, followed by the calculation of the dielectric constant from the response of polarization 
with respect to the field strength resulting in excellent agreement with the experimental value.
Moreover, we show that PNNP correctly 
predicts the field-induced red shift in the O-H stretching mode and the field-induced blue shift in the librational mode of liquid water, 
in very good agreement with results for ab-initio molecular dynamics. Perhaps the most surprising aspect of PNNP is that it predicts 
forces, energies and the dielectric response in excellent agreement with ab-initio data up to high electric fields, even though the PNNP 
was only trained on zero-field equilibrium structures. In {\it Discussion} we compare PNNP to previously introduced ML methods for 
calculation of molecular systems with electric fields and close our work in {\it Conclusions}.

\section*{Results}

\subsection*{Perturbed Neural Network Potential (PNNP)}
In our approach the interaction of the atomistic system with a homogeneous external electric field $\vec{E}$ is treated perturbatively 
via a 
\revchange{series}
expansion truncated at first order in the field~\cite{SzaboOstlund, zhang2020modelling, zhangsprik_dielectric,cox&sprik}
\begin{equation}
    \mathcal{H}_{\vec{E}}(\vec{r}^N,\vec{p}^N) = 
    \mathcal{H}_{0}(\vec{r}^N,\vec{p}^N) - 
    \vec{E} \cdot \vec{M}(\vec{r}^N)
    \, ,
    \label{eq:hamiltonian}
\end{equation}
where 
$\mathcal{H}_{0}(\vec{r}^N,\vec{p}^N)$ is the total unperturbed Hamiltonian comprised of the kinetic energy of the $N$ nuclei with momenta $\vec{p}^N$ and the electronic 
potential energy depending on all nuclear positions $\vec{r}^N$, $\mathcal{H}_{0}(\vec{r}^N,\vec{p}^N)\!=\!E_{\text{kin}}(\vec{p}^N)+ E_{\text{pot}}(\vec{r}^N)$,  
and 
$-\vec{E}\cdot \vec{M}(\vec{r}^N)$ is the perturbation induced by the electric field $\vec{E}$ acting on the total dipole moment of the system at \emph{zero field} $\vec{M}(\vec{r}^N)$.
The truncation to first order in the field is expected to be accurate for the weak and medium strong electric fields investigated 
in this work, as will be demonstrated further below. The field dependence of the dipole moment or, equivalently, higher order 
terms (polarizability, hyperpolarizability) may 
be added at stronger fields to account for the field-dependent perturbation of the electronic structure.\cite{resta2007theory}  
Applying Hamilton's equation of motion, we get for the force acting on atom $i$
\begin{equation}
    F_{i\xi} = 
    - \frac{\partial E_{\text{pot}}(\vec{r}^N)} {\partial r_{i\xi}} + 
    \sum_\zeta \frac{\partial M_\zeta}{\partial r_{i\xi}} E_\zeta
    \, ,
    \label{eq:force}
\end{equation}
where $\zeta\!=\!{x,y,z}$ and $\xi\!=\!{x,y,z}$ represent the three Cartesian coordinates and  $E_\zeta$ is the $\zeta$-component of $\vec{E}$. 
The first term is the force on the nuclei in the absence of an electric field and the second term is the field-induced contribution which can be written in terms of 
the transpose of the 
APT 
of atom $i$, 
$\mathcal{P}_i$\cite{Schienbein-2023-JCTC,Person-1974-JCP} (not to be confused with the polarization $\vec{P}$, Eq.~\ref{eq:polarization}), with elements  
\begin{equation}
    \frac{\partial M_\zeta}{\partial r_{i\xi}}
    \equiv 
     [ \mathcal{P}_i^\text{T} ]_{\xi\zeta}.  
    \label{eq:apt}
\end{equation}
In our approach we train 
ML models, 
one for the potential energy ($E_{\text{pot}}(\vec{r}^N)$) and one for the APT ($\mathcal{P}_i$) and use the corresponding 
forces, \eref{eq:force}, to carry out MD simulations in the presence of an external electric field. 
We emphasize that both quantities are trained {\it without} an external field present, in contrast to the schemes suggested 
before~\cite{Christensen-2019-JCP, Gastegger-2021-ChemSci, Gao-2022-NatCommun, Zhang-2023-NatCommun}. 
We use a committee~\cite{Schran-2020-JCP} of 2nd generation 
high-dimensional Neural Network potentials~\cite{Behler-2007-PRL, Behler-2021-ChemRev} (c-NNP) to model $E_{\text{pot}}(\vec{r}^N)$ and a  
E(3)-equivariant graph neural network to model the APT~(APTNN) as recently introduced by one of us~\cite{Schienbein-2023-JCTC}. 
For details of the force implementation we refer to section~{\it Methods}. In the following the combined c-NNP and APTNN model for the electronic potential 
energy including the field term, 
\revchange{$E_{\text{pot}}(\vec{r}^N)-\vec{E} \cdot \vec{M}(\vec{r}^N)$},
is simply referred to as 
\emph{``perturbed neural network potential''} (PNNP). 

PNNP MD simulations give access a number of important dielectric 
properties of solids, liquids and ionic solutions. The time derivative of the total dipole moment  
can be obtained by summing all APTs multiplied by the respective velocities of the nuclei, $v_{i\xi}$\cite{Schienbein-2023-JCTC} 
\begin{equation}
\dot{\vec{M}} \equiv  \frac{\text{d}\vec{M}}{\text{d}t}  = \sum_{i,\xi} \frac{\partial \vec{M}}{\partial r_{i\xi}} v_{i\xi} \; .
    \label{eq:dipole-time-derivative-from-apt}
\end{equation}
Thus, field-dependent IR spectra can be readily obtained from the
autocorrelation function of $\dot{\vec{M}}$ sampled along PNNP trajectories (see Eq.~\ref{eq:spectrum} below). 
Time-integration gives the total dipole moment of the cell,   
\begin{equation}
\vec{M} (t) = \vec{M} (t_0) + \int_{t_0}^t \text{d}t' \dot{\vec{M}}(t'), \label{eq:dipolemoment}
\end{equation}
and the polarization $\vec{P}$, 
 \begin{equation}
     \revchange{ \vec{P(t)} = \frac{\vec{M(t)}}{V} } ,
     \label{eq:polarization}
 \end{equation} 
where 
$V$ the volume of the simulation box.  
\revchange{In passing we refer to
the modern theory of polarization in solids\cite{resta2007theory}, where 
$\dot{\vec{M}}(t)/V$ is the transient current density that, when integrated over time 
gives the itinerant dipole moment.}
The polarization in Eq.~\ref{eq:polarization} is of major importance allowing 
for the calculation of relevant dielectric properties including 
\revchange{the dielectric constant (or ``relative permittivity'', }
see Eq.~\ref{eq:permittivity-field} below), capacitance and ionic conductivity.

\subsection*{Validation of PNNP}

\begin{figure}
    \centering
    \includegraphics[width=0.49\textwidth]{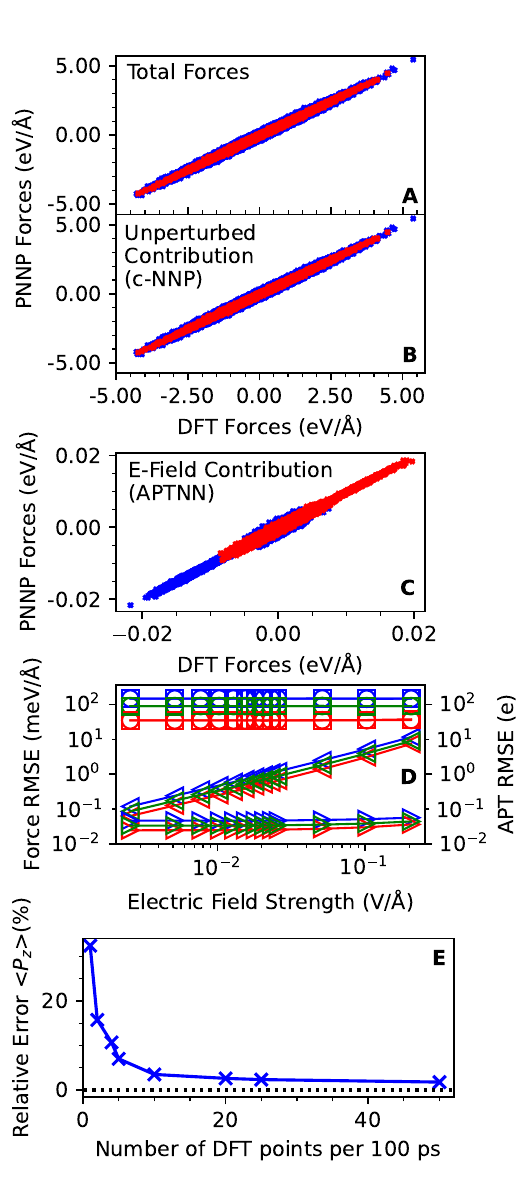}
    \caption{
        Error metrics of the trained PNNP at an exemplary field strength of \SI{0.0129}{\vpa}.
        Scatter plots are presented comparing the predicted PNNP force with the corresponding reference DFT forces for the total         
        PNNP force (A) and its individual components, namely the unperturbed force given by the c-NNP (B) and the field induced force obtained by the APTNN (C).
        Forces acting on O and H atoms are colored in blue and red, respectively.
        Mind the smaller scale in panel C since the field-induced forces are about two magnitudes smaller than the unperturbed forces.
        Panel D shows the overall RMSE as a function of the strength of the applied electric field, where the squares, circles, and left-facing triangles
        depict the force RMSE of the total PNNP force, the zero-field force, and the field-induced force, respectively.
        The right-facing triangles indicate the RMSE of the APT in units of elementary charge (right y axis).
        Mind the logarithmic scale and that the total PNNP force (squares) and the zero-field force (circles) largely overlaps.
        In E the relative error of the mean polarization in the direction of the field ($P_z$) obtained by integrating the total dipole moment time derivative (\eref{eq:dipolemoment}) 
        as a function of the sampling frequency of explicit DFT points, see text, is plotted.
    }
    \label{fig:rmse}
\end{figure}
   
We demonstrate our approach by simulation of pure liquid water at room temperature for different electric field strengths ranging 
from  
\num{0.0026} to \SI{0.2057}{\vpa}
(corresponding to \num{5e-5} and \num{4e-3} atomic units, respectively).
Recall, that the latter field strength is well in the regime of chemical bond activation~\cite{Hao-2022-NatCommun}.
Details on the training of the c-NNP and APTNN for liquid water are given in section {\it Methods}.  
Running a PNNP trajectory at an intermediate field strength of \SI{0.0129}{\vpa}, 
the drift in the conserved total energy 
is 
\num{-2.2e-9}
Hartree atom$^{-1}$ ps$^{-1}$
and
the mean magnitude of the center of mass momentum is 
\num{8.4e-9}
atomic units, 
see \fref{fig_si:conservation_plots} A and B, respectively.
These values  are representative for all performed PNNP simulation, i.e.\ all applied electric field strengths, 
and they are also 
typical for simulations without field  
attesting to the correct force implementation. 
Next we assess the quality of PNNP by validating 
the predicted forces along the trajectories against an independent test set unknown to the PNNP.  
At each field strength, 100~configurations are equidistantly extracted from 100~ps PNNP trajectories.
Then 
the total force on each nuclei is computed using DFT reference calculations at the same field strength and compared to the 
total force from PNNP (data shown in \fref{fig:rmse}A for an exemplary field strength of \SI{0.0129} {\vpa}).
We then decompose the DFT reference force into unperturbed and electric field-induced contributions by performing 
an additional DFT calculation on the same configurations, but without the external electric field.
The difference between the forces with and without the field isolates the field-induced forces.
This allows us to separately validate the unperturbed c-NNP force contribution, first term on the right-hand side of Eq.~\ref{eq:force},  
with the unperturbed DFT contribution (\fref{fig:rmse}B) and the electric field-induced APTNN contribution, second term on the 
right-hand side of Eq.~\ref{eq:force}, with the field-induced DFT contribution (\fref{fig:rmse}C). 

We find excellent correlation for PNNP and DFT total forces and its contributions. 
The 
root-mean-squared-error~(RMSE) in the atomic forces
as a function of field strength are shown in \fref{fig:rmse}D. 
The total force RMSEs are on the order of \SI{90}{\meva} and thus in line with previously published 
HD-NNP and c-NNPs on various systems~\cite{Natarajan-2016-PCCP, Schienbein-2022-PCCP, Schran-2020-JCP, Schran-2021-PNAS}.
Remarkably, we find that the RMSE for the unperturbed contribution remains almost constant at about \SI{90}{\meva} across all simulated field 
strengths up to \SI{0.2057}{\vpa}, although the training set is only composed of unperturbed, i.e., zero-field configurations and does not contain any polarized water 
samples. The field-induced force contributions are about two orders of magnitude smaller than the unperturbed contribution and so are their RMSEs. 
Notably, the \revchange{force} RMSE increases only in proportion with the field strength from \SI{8.6e-02} \meva at \SI{0.0026} {\vpa} to \SI{9.05} \meva at the largest 
simulated field of \SI{0.2057} {\vpa}. 
The reason for this is that the RMSE of the APT itself (\fref{fig:rmse}D, left y axis) remains close to constant across all simulated field strengths.
Due to error propagation (see \eref{eq:force}) that constant RMSE is amplified by the magnitude of the respective field strength.
Note that even at the highest field strength investigated, the force RMSE of the field contribution is about one magnitude smaller than the RMSE of the c-NNP.

Lastly, we validate the calculation of the total dipole moment along the PNNP MD trajectories as obtained by 
integration of the time derivative of the dipole moment according to Eq.~\ref{eq:dipolemoment}. We find that the dipole moment 
tracks the reference dipole moments obtained from explicit DFT calculations very well for about 10-100 ps depending on the field strength (Fig. S1C). 
However, at longer times deviations become large due to accumulation of errors when integrating over the finite time steps, even though $\dot{\vec{M}}(t)$ is accurately reproduced.
This problem is addressed here 
by calculating reference DFT dipole moments along PNNP MD trajectories in periodic intervals and integrating the 
time derivative of the dipole moment obtained from APTNN only from one DFT reference value to the next. Using this integration 
procedure we calculated the mean dipole moment averaged over the 100 ps PNNP MD trajectories, $\langle \vec{M}(t) \rangle$, 
as a function of the time interval between two DFT reference values. The relative errors with respect to the reference DFT mean dipole moments 
averaged over the same 100 ps PNNP MD trajectories
(calculated using a sampling frequency of \SI{1}{\per\pico\second}) are shown in \fref{fig:rmse}E. We find that the error decreases rapidly,  
to below 5\% for a time interval between two DFT reference 
dipole moments of 10~ps, for all field strengths. Hence, in practice only a limited number of additional DFT calculations are necessary 
to accurately calculate the dipole moment along the PNNP MD trajectories at a resolution that is only limited by the MD time step.
 
\begin{figure}[h]
    \centering
    \includegraphics[width=0.49\textwidth]{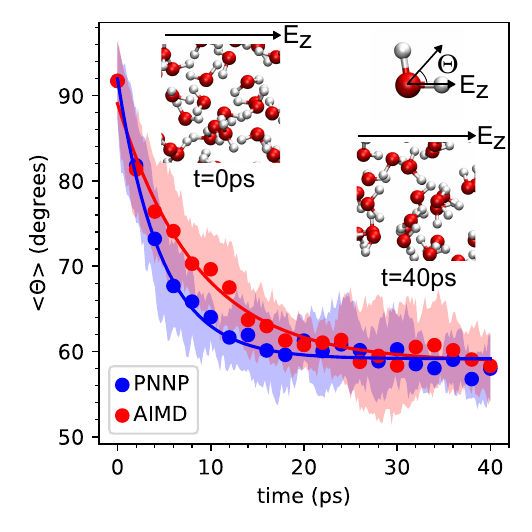}
    \caption{
        Orientational relaxation of water molecules in response to an applied electric field being switched on at $t=0$.
        The time evolution of the average water orientation angle $\langle\Theta\rangle$ comparing AIMD~(red) and PNNP simulations~(blue).
        The angle $\Theta$ is defined for each water molecule as the angle between the bisector of the two intramolecular OH bond vectors and the applied electric field vector $E_z$
        (as illustrated by the inset in the top right corner).
        The average $\langle\Theta\rangle$ is obtained by averaging $\Theta$ over all water molecules in a given configuration.
        The solid circles depict the average of $\langle\Theta\rangle$, while the shaded area indicates its corresponding standard deviation.
        The solid line represents an exponential fit to the solid circles.
        Note that \SI{90}{\degree} corresponds to a random orientation on average.
    }
    \label{fig:orientational-relaxation}
\end{figure}

\subsection*{Orientational relaxation dynamics}
Having validated the forces and polarization against DFT reference data, we now apply the PNNP to simulate the relaxation of water  
orientation in response to interaction with an electric field. 
We extract 10 independent configurations from a sample of liquid water equilibrated for 1~ns with field-free c-NNP MD and use them as starting 
configurations for 40~ps PNNP MD simulations, each run at a field strength of \SI{0.0257} {\vpa}. Additionally, we take the same 
initial configurations and perform explicit AIMD simulations at the same field strength. 
The orientation of the water molecules at a given time $t$ is described by the angle between the direction of the applied electric field and 
the 
\revchange{bisector between the two intramolecular OH bond vectors}
averaged over all water molecules, $\langle \Theta \rangle (t)$. 
The results are shown in  \fref{fig:orientational-relaxation}.  At $t\!=\!0$ the initial average orientation is \SI{90}{\degree} corresponding to the expectation value of the
randomly oriented water molecules
at equilibrium.
When the field is switched on at $t\!=\!0$ the relaxation dynamics obtained from PNNP MD is in excellent agreement with the results 
from AIMD for the first 5 ps and at longer times $> 20$ ps, both converging to a final average angle of \SI{58}{\degree}. 
Small deviations can be observed between 5-20~ps where PNNP MD gives a slightly faster exponential decay 
(time constant $\tau$ = 
4.66~ps$^{-1}$, $R^2\!=\!0.98$) 
than the data from AIMD ($\tau$ = 
8.53~ps$^{-1}$, $R^2\!=\!0.99$). 
However, since the error bars for the two methods overlap we ascribe this difference to statistical uncertainty.

\subsection*{Field Sweep}

\begin{figure}
    \centering
    \includegraphics[width=0.49\textwidth]{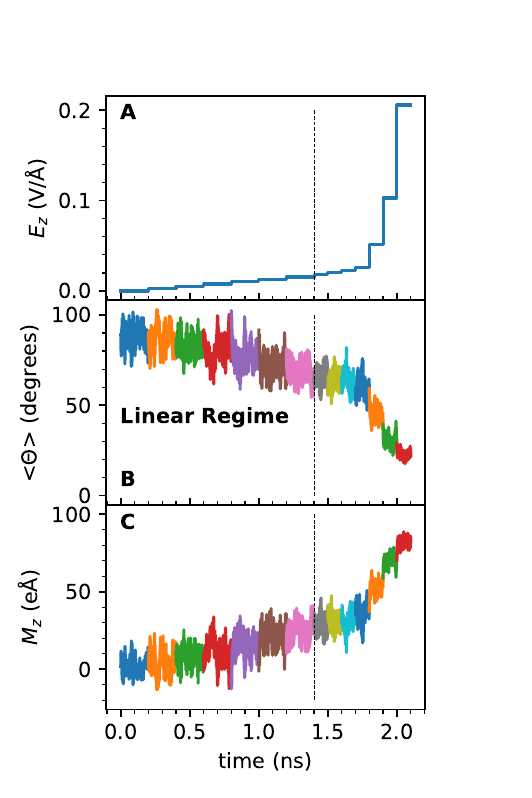}
    \caption{
        Progression of the performed field sweep MLMD simulation as a function of time.
        The magnitude of the applied electric field ($E_z$) is presented in panel A, 
        while the panels B and C show the average water orientation angle with respect the the field ($\Theta$, same as in \fref{fig:orientational-relaxation}) and the total dipole moment ($M_z$) in the direction of the field, respectively.
        The vertical dashed line at 1.4~ns indicates when the electric field strength exceeds \SI{0.0154}{\vpa} for the first time, thereby reaching the non-linear response regime. 
        Before 1.4~ns each field strength was kept constant for 200~ps, afterwards it was kept constant for 100~ps.
      }
    \label{fig:field-sweep}
\end{figure}

In the following we vary the strength of the field in time and monitor the structural and dielectric response of the water sample as obtained from PNNP simulations. 
The results are presented in \fref{fig:field-sweep}. The temporal profile of the applied electric field strength is shown in panel A, the average water orientation $\langle \Theta \rangle (t)$ 
in panel B and the total dipole moment in the direction of the applied electric field, $M_z(t)$, in panel C. 
We obtain an approximately linear response in the average orientation and a concomitant linear increase in the dipole moment 
as the field is stepped up from 0 to \SI{0.0154} {\vpa} (in increments of \SI{0.0026}{\vpa}). 
This is in line albeit somewhat lower than previously published estimates for the upper bound of the linear regime, 
\SI{0.03} to \SI{0.07}{\vpa}\cite{english2015perspectives,montenegro2021asymmetric}.
At larger field strengths 
the dielectric response becomes weaker indicating that the non-linear regime is reached. At \SI{0.2057}{\vpa} a strong orientational 
alignment along the field direction is observed, $\langle\Theta\rangle$\,=\,\SI{20}{\degree}. 
Notice that at these high field strengths 
the PNNP still predicts reasonably accurate forces (see \fref{fig:rmse}D) despite the absence 
of electronic polarization terms and
\revchange{any field-dependent training data}.
At \SI{0.4114}{\vpa}, 
\revchange{
we observe water splitting into a proton and a hydroxide ion.
}
It is well known that chemical bond activation starts to occur  at about \SI{0.1}{\vpa}~\cite{Hao-2022-NatCommun} 
and therefore 
this observation is not surprising.
While we could in principle further train the c-NNP to also include proton and hydroxide species, and thereby test even larger field strengths, accurate description of these species would require explicit inclusion of nuclear quantum effects~\cite{Ceriotti-2016-ChemRev}.
An additional set of simulations are run where the field strength is reversed from the value at the end of the linear regime, \SI{0.0154} {\vpa},  
down to 0 in increments of \SI{0.0026}{\vpa}. That sweep is detailed in \fref{fig_si:sweep}. We obtained very similar mean orientations and dipole moment 
as for the forward sweep demonstrating that the sample can be reversibly polarized and depolarized.   

\subsection*{Dielectric constant}

\begin{figure}[h!]
    \centering
    \includegraphics[width=0.49\textwidth]{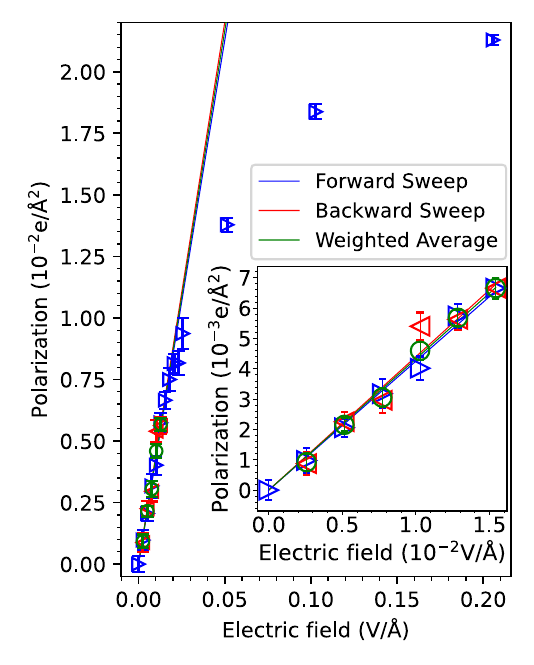}
    \caption{
        Average polarization as a function of the applied electric field strength.
        The blue right triangles depict the average polarization sampled from the field sweep PNNP MD simulation in \fref{fig:field-sweep} going from small to large fields (``forward-sweep'').
        Additionally, a backward sweep was performed (red left triangles) going from \SI{0.0154}{\vpa} (the end of the linear regime) back to zero field, see \fref{fig_si:sweep}.
        The green open circles mark the weighted average of the forward and backward sweep.
        The solid straight lines depict the linear fit to the respective data set (blue: forward sweep, red: backward sweep, green: weighted average), considering the linear response regime only ($E \leq \SI{0.0154}{\vpa}$), from which the dielectric constant is calculated (see \eref{eq:permittivity-field}).
        The inset contains an amplification of the linear response regime, showing the very same data and using the same color code; mind the smaller scales on both axes.
    }
    \label{fig:pol-vs-field}
\end{figure}

The time series of dipole moments in \fref{fig:field-sweep}C are averaged 
\revchange{individually for each applied electric field strength}
to obtain the polarization of the water sample along the field direction, $P_z$ (Eq.~\ref{eq:polarization}), 
as a function of the field strength. 
The results are shown in \fref{fig:pol-vs-field} for the forward and backward sweeps (data in blue and red, respectively). The data points in 
the linear regime are shown magnified in the inset of  \fref{fig:pol-vs-field}. They are fit to straight lines and the slopes used to obtain the static 
dielectric constant, $\epsilon_r$, according to Eq.~\ref{eq:permittivity-field}, 
\begin{equation}
    \epsilon_r = 1 + \frac{1}{\epsilon_0} 
    \revchange{\frac{\partial \langle P_z\rangle}{\partial E_z}} 
    \; ,
    \label{eq:permittivity-field}
\end{equation}
where $\epsilon_0$ is the vacuum permittivity. 
We obtain values 
\revchange{
$\epsilon_r\!=\!77.9 \pm 2.7$ ($R^2\!=\!0.99$), $80.6 \pm 2.7$ ($R^2\!=\!0.96$) for forward and backward sweep
and $\epsilon_r\!=\!79.3 \pm 2.2$ ($R^2\!=\!0.99$, data in green) 
}
from a linear fit to the weighted average data for forward and backward sweep, in  
excellent agreement with the experimental value of $\epsilon_r\!=\!78.4$~\cite{Fernandez-1997-JPhysChemRefData}. 
Similar values would be obtained using only one data point due to the excellent linear correlation. 
Remarkably, the dielectric constant was converged after only about 175ps of simulation time \revchange{per applied electric field} (see \fref{fig_si:epsilon-convergence-field}), similar to what was previously reported for 
classical MD simulations employing finite field Hamiltonians\cite{zhangsprik_dielectric}. This is an order of magnitude less simulation time than what is 
required to calculate the dielectric constant from the fluctuations of the polarization~\cite{kirkwood1939dielectric,NEUMANN_dielectric,DPS_ewald_dielectric,zhangsprik_dielectric}, 
\begin{equation}
    \epsilon_r = \epsilon_\infty + \frac{1}{3\epsilon_0 k_\text{B}VT} 
    \left(
    \langle \vec{M}^2 \rangle
    -
    \langle \vec{M} \rangle^2
    \right) \, ,
    \label{eq:permittivity-variance}
\end{equation}
where the total dipole moment, $\vec{M}$, is sampled at zero electric field 
and $\epsilon_\infty\!=\!1.72$~\cite{galli_dielectric} is the optical dielectric constant for liquid water. 
Indeed, sampling the total dipole moment $\vec{M}$ along zero field c-NNP trajectories 
we obtain a converged value $\epsilon_r\!=\!90.6 \pm 1.7$ only after 3 ns (see \fref{fig_si:epsilon-convergence-variance})
similarly as in previous simulations~\cite{howvdwdeterime_water_properties}.
The reason for the difference in the dielectric constant from the field sweep and from zero field 
simulation is not known but could be due to a number of reasons. The fluctuations of the polarization are likely to be more sensitive 
on simulation details than the mean values, e.g., thermostat used, finite system size and remaining inaccuracies of the PNNP.  
Moreover, the approximations made to derive the dielectric constant in terms of the polarization fluctuations from its basic definition 
\eref{eq:permittivity-field}~\cite{Neumann_second_order} could also contribute to the difference.  
       
\subsection*{Field-dependent IR spectra}

\begin{figure}[h]
    \centering
    \includegraphics[width=0.49\textwidth]{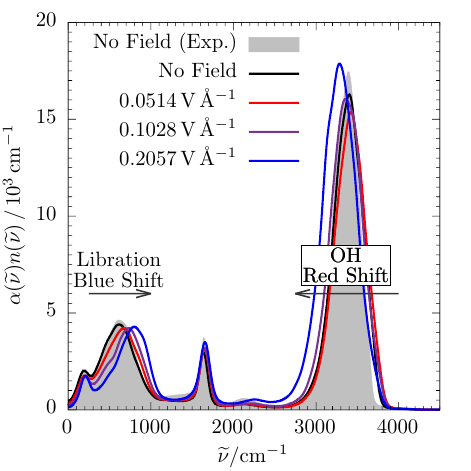}
    \caption{
        Field-dependent Beer-Lambert absorption coefficients of liquid water.
        The Beer-Lambert absorption coefficient times the frequency dependent refractive index $\alpha(\waveno)n(\waveno)$ of room temperature liquid water is presented as a function of the applied electric field strength.
        The black solid line depicts the spectrum of water without the presence of any field, and the red, purple, and blue curves depict the corresponding spectra for field strengths of 
        \num{0.0514}, \num{0.1028}, and \SI{0.2057}{\vpa}, respectively.
        The experimental spectrum at zero field is taken from Ref.~\cite{Bertie-1996-ApplSpectrosc} and shown for reference as shaded grey area.
    }
    \label{fig:spectra}
\end{figure}

The APTs and velocities along the PNNP trajectories give direct access to the time derivative of the dipole moment, $\dot{\vec{M}}$, according to Eq.~\ref{eq:dipole-time-derivative-from-apt}.  
The autocorrelation function of $\dot{\vec{M}}$ gives the frequency ($\omega$)-dependent Beer-Lambert absorption coefficient of IR spectroscopy, $\alpha(\omega)$, 
\begin{equation}
    \alpha(\omega)  n(\omega)
    = \frac{\pi}{3Vc\epsilon_0 k_\text{B} T} 
    \frac{1}{2\pi} 
    \int_{-\infty}^\infty dt \,
    e^{-i\omega t}
    \left<\dot{\vec{M}}(0) \dot{\vec{M}}(t) \right>, 
    \label{eq:spectrum}
\end{equation}
where $n(\omega)$ is the frequency-dependent refractive index, $c$ is the speed of light in vacuum, $k_\text{B}$ the Boltzmann constant and 
$T$ the temperature.
In \fref{fig:spectra} we present the calculated 
\revchange{$\alpha(\omega)n(\omega)$}
as obtained from c-NNP simulations at zero field 
and from PNNP trajectories at different field strengths
\revchange{in the non-linear response regime}.
The experimental IR spectrum at zero-field~\cite{Bertie-1996-ApplSpectrosc}  
is very well reproduced using the RPBE-D3 functional, as reported previously~\cite{Schienbein-2023-JCTC}.
The spectrum remains remarkably insensitive to the presence of an electric field in the linear regime and beyond, up to about \SI{0.0514}{\vpa}. 
For larger fields we observe a systematic red-shift of the intramolecular OH stretching vibration at around \SI{3500}{\invcm}, 
by approximately \SI{100}{\invcm} at \SI{0.2057}{\vpa}.  
A red-shift of the \ce{OH} stretch is generally ascribed to formation of stronger intermolecular hydrogen bonds~\cite{Rey-2002-JPCA, Lawrence-2003-JCP, Fecko-Science-2003}, 
here induced by the external electric field. This notion is further supported by the systematic blue shift of the librational band at around \SI{3500}{\invcm} by 
approximately \SI{200}{\invcm} at \SI{0.2057}{\vpa} suggesting that the interaction with the field leads to a stiffening of the potential for rotational motion of the water 
molecules~\cite{Schienbein-2020-ANIE}. This is in line with a previous study describing water at these field strengths to be more ``ice-like''~\cite{Cassone-2019-PCCP}.  
The trends in the IR spectra agree very well with previously published results obtained from other approaches that differ in many aspects from our method, 
AIMD simulations with external electric fields~\cite{Cassone-2019-PCCP} and MLMD simulations using the FIREANN-wF model~\cite{Zhang-2023-NatCommun}.

\section*{Discussion}
In this work we extend MLMD simulations to include the interaction with an external electric field by adding the 
\revchange{field induced perturbation up to first-order}
to the unperturbed Hamiltonian \eref{eq:hamiltonian}.
The resulting force equation (\eref{eq:force}) allows us to separately calculate the unperturbed forces from a standard MLP~(here: a c-NNP)  
and the field-induced forces from the APTNN. The force calculation therefore rigorously follows from the laws of electrostatics without any 
additional approximations. This has several advantages:
First, the approach is modular because the electric field contribution is independent from the employed unperturbed potential energy surface.
As such, the APTNN can be coupled with any MLP (not only NNPs) or even with force fields to include the interaction with an external electric field.
The level of theory of the reference electronic structure calculations for the training of APTNN and MLP can be chosen differently in accord with the  
accuracy requirements for unperturbed and perturbed potentials. Moreover, the modular approach allows one to assess the accuracy of each component 
separately.  
Second, APTs are well defined for any atomistic system and thus APTNNs can be trained ``out of the box", without 
requiring any conceptual tailoring or adjustment or the use of arbitrarily defined proxies (e.g. atomic charges, molecular 
dipole moments, or similar)\cite{Schienbein-2023-JCTC}. Importantly, APTs are uniquely defined when using periodic boundary conditions 
as they quantify a {\it change} of the total dipole moment or polarization. This is in contrast to the total dipole moment, which is multi-valued in 
periodic boundary conditions~\cite{resta_MTP_1992}.
Third, the method can be systematically improved by including higher order terms 
\revchange{to the perturbation}, e.g.\ polarizability and 
hyperpolarizability~\cite{SzaboOstlund}.
This would allow for more accurate simulations at
\revchange{even larger field strengths as investigated herein}.
\revchange{Through the inclusion of the next higher term in the multipole expansion~\cite{jackson-classical-1999}, one could potentially also describe electric field gradients}.
\revchange{Finally, the approach does not require training as a function of the applied electric field, even up to rather strong fields at the onset of chemical bond activation.}

These features make our method distinct from other approaches recently introduced to model the interaction with external 
electric fields in ML simulations.
\revchange{
    Field-dependent training data is usually required because the electric field is direcly embedded into a single MLP~\cite{Christensen-2019-JCP,Gastegger-2021-ChemSci, Gao-2022-NatCommun,Zhang-2023-NatCommun}.
}
\revchange{For example, to perform finite-field simulations on  liquid water, the FIREANN-wF model trained atomistic forces in the presence of an electric field~\cite{Zhang-2023-NatCommun}}.
The model was then successfully used to calculate response functions, such as dipole moments and polarizabilities, and from these 
quantities field-dependent vibrational spectra of liquid water~\cite{Zhang-2023-NatCommun}. 
However, the field-induced contribution to the total energy and forces is thus only
\emph{``learned implicitly in the force-only training''}~\cite{Zhang-2023-NatCommun}.
In contrast to our method, this approach is not modular and it 
is not obvious how it can be systematically improved. 
Earlier approaches~\cite{Christensen-2019-JCP, Gastegger-2021-ChemSci} rely on the definition of proxies, 
such as fictitious partial charges\cite{Christensen-2019-JCP} or atomistic dipole moments\cite{Gastegger-2021-ChemSci}.
\revchange{It also does not depend on the definition of molecular reference frames, which is e.g.\ required to train the position of Wannier centers~\cite{Gao-2022-NatCommun, Cools-2022-JCTC}.}
The APTNN in turn, which is entirely responsible for the field-induced atomic forces, trains on a well-defined physical 
observable~\cite{Person-1974-JCP, Schienbein-2023-JCTC} such that it is guaranteed to be meaningful independent of the system of interest.
Finally, our method exclusively trains on unperturbed, zero-field data, the electric field dependence is therefore offloaded to the MD sampling 
by \eref{eq:force}. As such, the electric field is not an input parameter of the 
ML model.

We performed PNNP simulations at several different field strengths and compared our results against reference AIMD calculations
and experimental literature data. In general, we found very good to excellent agreement for all properties investigated. We demonstrated reversible orientational 
polarization and depolarization of the water molecules as the electric field is stepped up and down, with orientational relaxation times 
that are in good agreement with AIMD data (\fref{fig:orientational-relaxation}). This permitted successful calculation of 
the relative permittivity, $\epsilon_r$, on simulation time scales that are an order of magnitude shorter than in more standard approaches 
that suffer from the very slow convergence of the polarization fluctuations at zero field. We obtained a value 
\revchange{$\epsilon_r = 79.3 \pm 2.2$}
in excellent agreement with experiment, $\epsilon_r = 78.41$ at 298.14~K and 1~atm~\cite{Fernandez-1997-JPhysChemRefData}  
(\fref{fig:pol-vs-field}). 
By definition (\eref{eq:permittivity-field}), the dielectric constant is a direct measure of a system's response to an applied electric field.
The observed, almost quantitative, agreement with previous simulations and the experiment therefore validates our approach in the linear response regime below \SI{0.0154}{\vpa}.
With respect to the non-linear regime, we calculated IR spectra and compared them with previous explicit AIMD simulations~\cite{Cassone-2019-PCCP}.
The systematic peak shifts in the IR spectrum for libration and intramolecular OH stretch vibration at high electric fields match 
\revchange{that reference data very well, thereby validating our approach in the non-linear response regime as well}.
 
The almost quantitative agreement obtained for the dielectric constant is remarkable, because one generally cannot expect perfect agreement 
between electronic structure calculations at the level of DFT and experiments. Moreover, nuclear quantum effects~\cite{Ceriotti-2016-ChemRev} were ignored in our simulations. The 
RPBE-D3 functional in combination with classical MD simulation is known to describe water, aqueous solutions and interfaces over 
a wide range of temperatures and pressures very well~\cite{Schienbein-2020-ANIE, Schienbein-2018-JPCB, Schienbein-2020-PCCP,Schienbein-2023-JCTC, 
howvdwdeterime_water_properties,Imoto-2015-PCCP, ForsterTonigold-JCP-2014, Gross-2022-ChemRev}.
With the help of PNNP simulations we could show that the excellent performance of the RPBE-D3 functional in describing radial distribution functions,
\revchange{self-diffusion, orientational relaxation processes},
and vibrational spectra of liquid water
\revchange{with respect to experimental data},
extends to its dielectric response properties. 
This is in part because remaining deficiencies of this functional tend to 
be effectively compensated by missing nuclear quantum effects. 
Indeed, 
\revchange{our computed IR spectrum at zero-field is in almost quantitative agreement with experimental data. Moreover, }
our computed peak shifts in the IR spectra 
\revchange{in the presence of an applied electric field} 
are in almost 
quantitative agreement with the results reported in the MLMD simulations of Ref.~\cite{Zhang-2023-NatCommun}.
\revchange{Therein} 
a more accurate and computationally expensive revPBE0-D3 hybrid functional was employed in combination with path integral 
MD simulations~\cite{Zhang-2023-NatCommun} to explicitly account for nuclear quantum effects.

We emphasize that in our method the machine learning models (c-NNP, APTNN) are exclusively trained on unperturbed, zero-field configurations 
and properties. Polarized water structures were not included in the training. Thus it is highly remarkable that the trained 
c-NNP and APTNN faithfully model ``unseen'' polarized structures up to field strengths of about \SI{0.21}{\vpa}. 
It also shows that inclusion of 
higher order terms in the perturbation series expansion, i.e.\ polarizabilities or hyperpolarizabilities,
is not necessary at least up to this field strength, as far as the accuracy of nuclear forces in 
PNNP MD simulations is concerned. 
At a field strength of about \SI{0.41}{\vpa} we observed the onset of water 
dissociation into \ce{H+} and \ce{OH-}, causing the c-NNP to extrapolate and, ultimately, to fail. Our simple approach 
therefore covers a quite large regime of electric field strengths, from smaller fields in the linear regime up to chemical bond 
activation around \SI{0.1}{\vpa}\cite{Hao-2022-NatCommun}. 

\section*{Conclusions}
In conclusion, we have implemented a ML methodology to run MD simulations in the presence of an 
external homogeneous electric field. Key to this development is the APT, which is trained by a ML 
model and used to compute the field induced forces. The latter are combined with the forces obtained 
from a standard MLP describing the interatomic potential at zero-field.
The method is rigorous, approximation-free, and does not require the definition of artificial proxies as target values for training, 
because the atomic polar tensor is a well-defined physical observable, even in periodic boundary conditions.
By validating our approach against explicit AIMD simulations and experimental and computational data from the literature, 
we demonstrate that it accurately models liquid water in external homogeneous electric fields up to 
field strengths of about \SI{0.21}{\vpa}. This is deep in the non-linear response regime and at the onset for 
chemical bond activation. 
This is possible by training the MLP on zero-field configurations only.
We expect that our method will be useful for ab-initio-level calculation of other field-dependent properties of 
condensed phase systems not covered in this work including ionic conductivity and capacitance.    

\section*{Methods}
\subsection*{Implementation of PNNP}
The unperturbed potential energy, $E_{\text{pot}}$, and the corresponding forces, first term on the right-hand side
of \eref{eq:force} are modelled by a committee
of 2nd generation high-dimensional Neural Network potentials,\cite{Behler-2007-PRL, Behler-2021-ChemRev}
as recently implemented in the \texttt{cp2k} software package.\cite{Schran-2020-JCP}  
The field dependent force contribution related to the APT, second term on the right-hand side of \eref{eq:force}, 
is modelled by an APTNN, which is based on the E(3)-equivariant graph neural network \texttt{e3nn}\cite{e3nn} 
based on  the \texttt{PyTorch} library\cite{pytorch}. A new force evaluation environment was added to 
\texttt{cp2k} linking the fortran-based \texttt{cp2k} and the python/C++ based \texttt{PyTorch} in a 
client/server approach. Inspired by the i-PI implementation\cite{kapil2019pi} \texttt{cp2k} launches and connects to a 
python server that waits to receive configurations and sends back APTs predicted by the APTNN model.
The APTs are then used in \texttt{cp2k},
to evaluate the corresponding force contributions. 
They are 
added to the c-NNP forces
\revchange{using the built-in \emph{mixing} force environment}
to obtain the total forces for propagation of the atoms using the velocity-Verlet algorithm.  

\subsection*{Training of c-NNP and APTNN}
The committee members of the c-NNP were trained using \texttt{n2p2}\cite{singraber2019parallel}.
The parameters for the neural network were taken from a previous MLP study on 
liquid water\cite{howvdwdeterime_water_properties}, in conjunction with generic symmetry 
functions\cite{Schran-2020-JCP}. The c-NNP is trained on the energies and forces obtained from 
DFT calculations at the level of RPBE-D3 as detailed further below using the active learning procedure reported in 
Ref.~\cite{Schran-2020-JCP}. With regard to the APT, we use the recently published APTNN developed 
by one us\cite{Schienbein-2023-JCTC} containing 10,368 individual APTs in its training set at the 
level of RPBE-D3~\cite{Schienbein-2023-JCTC}. This network has previously been benchmarked, 
and was shown to accurately reproduce the IR spectrum of liquid water under ambient conditions. 

\subsection*{Reference DFT calculations}
Electronic structure calculations for generation of reference training and test data were performed 
using \texttt{cp2k}\cite{cp2k_code} version 2023.1 and the quickstep  module~\cite{cp2kquickstep}.
The RPBE functional was evaluated by the \texttt{libxc} package~\cite{Marques-ComputPhysCommun-2012}
and supplemented by D3 dispersion corrections~\cite{d3correction}.
We employed a mixed Gaussian orbital/plane wave (GPW) basis set~\cite{Lippert-MolPhys-1997} with 
a plane wave cutoff of 600 Ry and a relative cutoff of 20~Ry, and the Gaussian orbitals are constructed 
using the triple-$\zeta$ quality TZV2P basis set\cite{VandeVondele-2007-JCP} including polarization functions
as employed previously\cite{Imoto-2015-PCCP, Schienbein-2018-JPCB, Schienbein-2020-PCCP, Schienbein-2020-ANIE}.
Core electrons are described by norm-conserving Goedecker-Teter-Hutter (GTH) pseudopotentials~\cite{gthpseudopot, relativisticpseudopot}.
Homogeneous electric fields were treated by the approach introduced by Umari and Pasquarello~\cite{Umari_Pasquarrello} as implemented in \texttt{cp2k}.
Dipole moments of the simulation cell ($\vec{M}$) were obtained from the DFT calculations via maximally localized Wannier functions~\cite{Marx-Book-2009}.

\subsection*{Charge conservation}

Charge conservation is 
enforced by the acoustic sum rule 
\begin{equation}
    \sum_i \frac{\partial \mu_{\alpha}}{\partial R_{\beta}^i} = 0
\end{equation}
for the atomic polar tensors~\cite{acousticsum_og,apt_gaussian_planewaves_calc,acoustic_sum_rule_paper}.
This is done by calculating the sum of every component divided by the number of atoms and distributing any excess evenly across all atoms in the system. 
Note that this correction is akin to correcting atomic charges, as it has previously been done in the literature, distributing the excess charge evenly across the total system~\cite{unke2019physnet,unke2021spookynet,song2024charge}.

\subsection*{Simulation details}
All calculations including training, testing and production runs were carried out for a 128 water molecule box of 
length \SI{15.6627}{\angstrom} employing periodic boundary conditions (density \SI{0.996}{\kilo\gram\per\liter}).
All finite field PNNP MD simulations were carried out in the NVT ensemble using a CSVR thermostat with a time constant of 1 ps~\cite{bussi2007canonical} 
and a MD integration time step of 1 fs. 
Zero field c-NNP MD simulations were carried out in the NVT
ensemble using a Nose-Hoover thermostat\cite{nose1984unified,nose1984molecular} and a time step of 0.5 fs. 

To illustrate the reversability of the electric field within a single PNNP MD simulation, we performed a field sweep starting at zero field up to a field strength of \SI{0.0154}{\vpa} (the end of the linear regime).
Therein, the field strength was increased every 200~ps by \SI{0.0026}{\vpa}.
Additionally, we then performed a backward sweep back to zero field by decreasing the applied electric field strength by \SI{0.0026}{\vpa} at every 200~ps as illustrated in \fref{fig_si:sweep} in the SI.
The average water orientation angle $\langle\Theta\rangle$ (see main text) and the polarization in the direction of the field are plotted as well.
From the slope of the polarization as a function of the applied field strength the dielectric constant can be estimated (see \eref{eq:permittivity-field} in the main text).
The field sweep PNNP MD can be used to calculate the dielectric constant twice, namely from the forward and from the backward sweep.

To calculate the Beer-Lambert absorption coefficients in \fref{fig:spectra}, we follow a previous simulation protocol~\cite{Schienbein-2020-ANIE}:
For each electric field strength (\num{0.0514}, \num{0.1028}, and \SI{0.2057}{\vpa}) 20 independent configurations were sampled from the field sweep PNNP simulation in \fref{fig:field-sweep}.
For each of these configurations short (20~ps) PNNP simulations in the NVE ensemble were conducted. 
The spectra are calculated for each of the 20 simulations independently and their average is shown in \fref{fig:spectra}.

\subsection*{Convergence of the dielectric constant}

The dielectric constant can either be calculated from the variance of the total dipole moment without any electric field present (\eref{eq:permittivity-variance}) or from the response of the system to an applied electric field (\eref{eq:permittivity-field}).
We first assess the convergence of the dielectric constant using the variance at zero field as shown in \fref{fig_si:epsilon-convergence-variance} in the SI.
We observe that the variance is converged after simulating for about 3~ns in total (panel C), while sampling snapshots from the simulation every pico second (panel B).
We also assess the convergence of the dielectric constant as a function of time using the change in polarization caused by an applied electric field (see \eref{eq:permittivity-field} in the main text).
To do so, we separately compute the mean polarization of the forward and backward field sweep up to time $t$ and then perform a linear fit in full analogy to \fref{fig:pol-vs-field} yielding a dielectric constant for the forward and the backward sweep, respectively.
The obtained dielectric constant as a function of time is presented in \fref{fig_si:epsilon-convergence-field}.
We find that after about 175~ps, the value of the dielectric constant is converged.
Each applied external field strength should thus be simulated for about 150~ps.
Herein we run the simulation at each field strength for 200~ps (see above), taking equilibration to the new field strength into account.
Due to the almost perfect linear fit of the polarization with respect to the applied field strength (see \fref{fig:pol-vs-field}), only a few field simulations are practically required and thus the computational cost is thus significiantly lower compared to converging the variance of the dipole moment at zero field (see \fref{fig_si:epsilon-convergence-variance}).

\section*{Data Availability}
The full data for this study totals more than 1 TB, so it is in cold storage accessible by the corresponding authors 
and available upon reasonable request.

\section*{Code Availability}
All field simulations were performed using the \texttt{cp2k} software package, that has been customised in-house and will be
made available in a GitHub repository linked in the supporting information.
\revchange{
    The atomic polar tensor neural network and all scripts used to train and predict atomic polar tensors is publicly available on:
    https://github.com/pschienbein/AtomicPolarTensor.
}

\printbibliography

\section*{Acknowledgements}
K.J. gratefully acknowledges a PhD studentship co-sponsored by University College London and Pacific Northwest National Laboratory (PNNL) 
through its BES Geosciences program supported by the U.S. Department of Energy's Office of Science, Office of Basic Energy Sciences, 
Chemical Sciences, Geosciences and Biosciences Division. 
This work was further supported by an individual postdoc grant to P.S.\ funded by 
the Deutsche Forschungsgemeinschaft (DFG, German Research Foundation) under project number 519139248 (Walter Benjamin Programme).
J.B. and P.S acknowledge EPSRC - UKRI for award of computing time via a ARCHER2 Pioneer Project (ARCHER2 PR17125).   
Via our membership of the UK's HEC Materials Chemistry Consortium, which is funded by EPSRC (EP/L000202, EP/R029431), 
this work also used the ARCHER UK National Supercomputing Service (http://www.archer.ac.uk). 

\section*{Author Contributions}
K.J. implemented PNNP under the supervision of P.S., K.J. carried out and analyzed all simulations. P.S. calculated the IR spectra
and helped analyzing the data. P.S and J.B designed the research. K.J., P.S. and J.B wrote the manuscript, with input from K.R. 
All authors reviewed and discussed the manuscript.

\section*{Competing Interests}
The authors declare no competing interests.

\section*{Additional Information}
Supplementary information. The online version contains supplementary material available.

\end{document}


\maketitle

\section*{Energy, Charge and  Momentum conservation}
\begin{figure}[b!]
    \centering
    \includegraphics[width=0.49\textwidth]{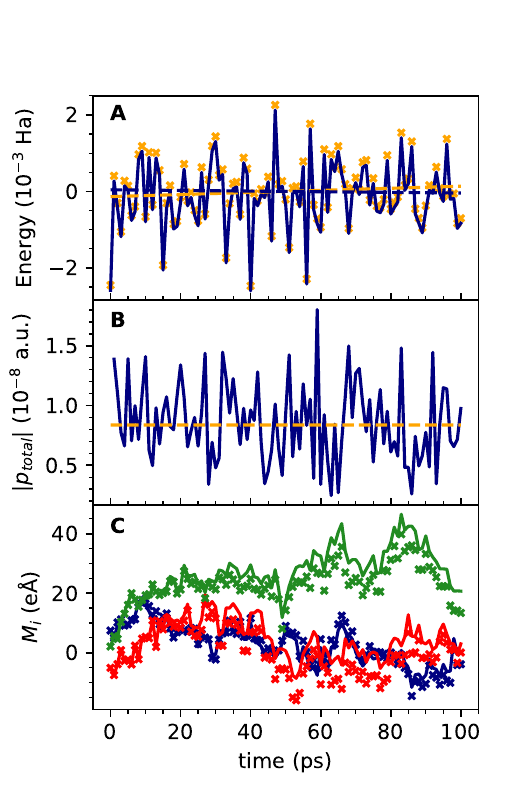}
    \caption{
        Panel A shows the difference between the field-induced energy contribution and its mean value as a function of simulation time calculated from the integrated polarization (solid blue line) and from DFT reference calculations (yellow crosses).
        The dashed lines illustrate linear fits to that data using the same color code.
        Panel B shows the magnitude of the total momentum as a function of simulation time.
        Panel C presents the polarization as a function of simulation time, calculated by integration (\eref{eq:dipolemoment}, solid lines) and from explicit DFT reference calculations (crosses)
        for the three spatial coordinatex x (blue), y (red), and z (green).
        Mind that the electric field is applied along the z direction.
    }
    \label{fig_si:conservation_plots}
\end{figure}

\clearpage
\section*{Field Sweep}
\begin{figure}[h]
    \centering
    \includegraphics[width=0.49\textwidth]{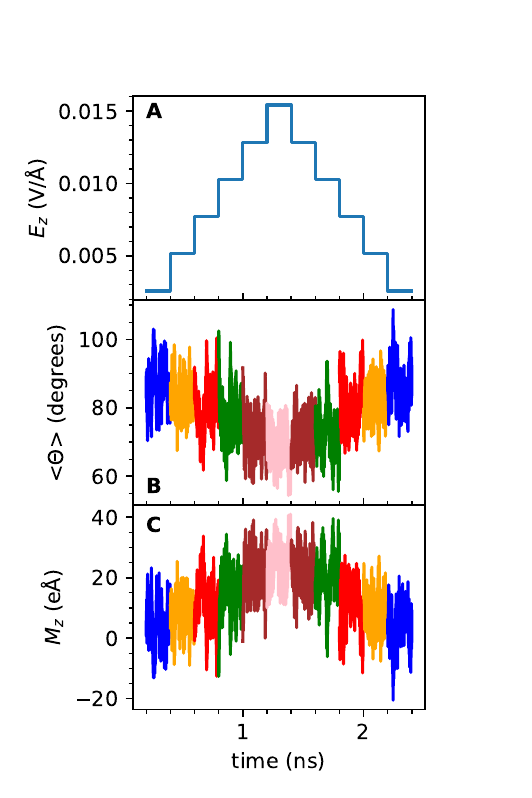}
    \caption{
        PNNP MD simulating  a field sweep going from 0 to \SI{0.0154}{\vpa}, by increasing the applied field strengths by \SI{0.0026}{\vpa} every 200~ps.
        The applied field strength is depicted in panel A, while the response of the water molecules in terms of the average water orientation $\langle\Theta\rangle$ (see main text) and of the polarization along the field ($M_z$) is shown in panels B and C, respectively.
    }
    \label{fig_si:sweep}
\end{figure}

\clearpage
\section*{Convergence of dielectric constant}
\begin{figure}[h!]
    \centering
    \includegraphics[width=0.49\textwidth]{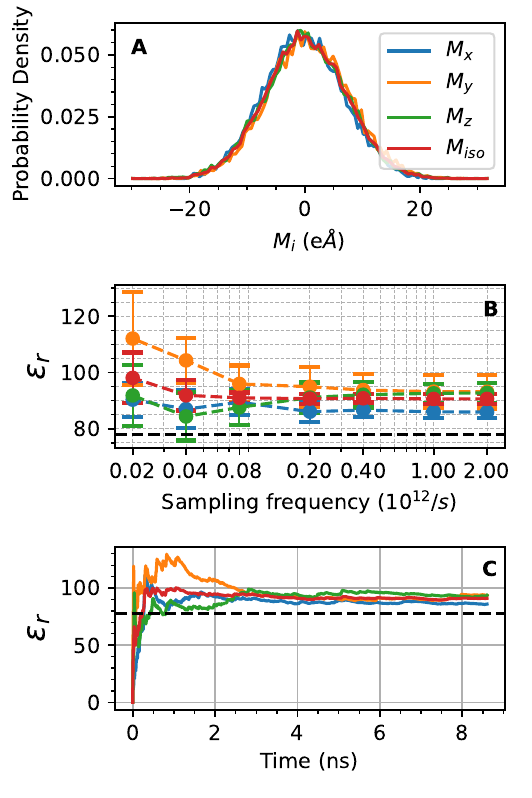}
    \caption{
        Convergence of the dielectric constant as a function of time using the variance of the total dipole moment along a zero field MD simulation, see \eref{eq:permittivity-variance} in the main text.
        Panel A depicts the ensemble distribution function of the total dipole moment individually for the different cartesian components and their average ($M_\text{iso}$, red).
        The convergence of the dielectric constant is presented as a function of the sampling frequency (panel B) and the total simulation time (C), both following the same color code as in panel A.
    }
    \label{fig_si:epsilon-convergence-variance}
\end{figure}
\begin{figure}[h!]
    \centering
    \includegraphics[width=0.49\textwidth]{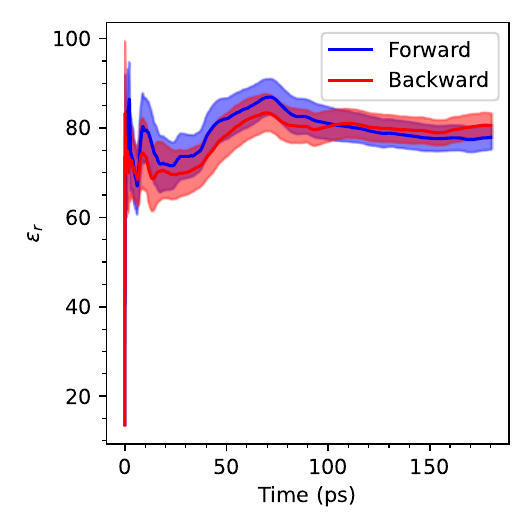}
    \caption{
        Convergence of the dielectric constant as a function of time during the forward (blue) and backward (red) electric sweep PNNP MD simulation depicted in \fref{fig_si:sweep} using the change in polarization caused by an applied electric field \eref{eq:permittivity-field}.
        The shaded area illustrates the corresponding standard deviation of the mean.
    }
    \label{fig_si:epsilon-convergence-field}
\end{figure}

